\documentclass{elsart}
\usepackage[dvips]{graphicx}
\begin{document}

\newcommand{\beq}{\begin{equation}}
\newcommand{\eeq}{\end{equation}}
\newcommand{\bea}{\begin{eqnarray}}
\newcommand{\eea}{\end{eqnarray}}
\begin{frontmatter}
\title{Volatility Driven Market in a Generalised Lotka Voltera Formalism}
\author{Yoram Louzoun}

\address{Department of Molecular Biology, Princeton University, Princeton, NJ 08540,email: ylouzoun@princeton.edu}

\author{Sorin Solomon}
\address{Racah institute of physics, Givat Ram Hebrew University, Jerusalem 91904 Israel.}
\date{\today}

Pacs:89.65.Gh 
\begin{keyword}
volatility auto-correlations, power laws, econophysics,
Lotka Volterra, stochastic logistic, behavioral finance
\end{keyword}
\begin{abstract}
The Generalized Lotka Voltera (GLV) formalism has been introduced in
order to explain the power law distributions in the individual wealth
($w_i (t)$) (Pareto law) and financial markets returns (fluctuations)
($r$) as a result of the auto-catalytic (multiplicative random)
character of the individual capital dynamics.

As long as the multiplicative random factor ($\lambda$) is extracted
from the same probability distribution for all the individuals, the
exponent of the power laws turns out to be independent on the time
variations of the average ($< \lambda >$). This explains also the
stability over the past century of experimentally measured Pareto
exponent.

In contrast to the scaling properties of the single time
("unconditional") probability distributions, the (auto-)correlations
between observables measured at different times are not correctly
reproduced by the original GLV, if the variance (${\sigma}^2$) of
$\lambda$ is time independent.  In the GLV formalism the volatility
($r^2$) auto-correlations decay exponentially while the measurements
in real markets indicate a power law with a very small exponent.

We show in the present paper that by making the variance of the
individual wealth changes ${\sigma}^2$ a function of the market
volatility $<r^2>$, one correctly reproduces the market volatility
long range correlations.

Moreover, we show that this non-trivial feedback loop between the
market price volatility and the variance of the investors wealth leads
to non-trivial patterns in the overall market trends. If the feedback
is too strong, it may even endanger the market stability.
\end{abstract}
\end{frontmatter}

\newpage

\section{Introduction}

In the last decade, the Microscopic Representation techniques
 were used in a wide range of subjects\cite{mr}. In particular, Levy
Levy and Solomon \cite{lls,lls1}(LLS) have devised a model of the
financial markets in terms of a large number of virtual investors
characterized each by a current wealth, portfolio structure,
probability expectations and risk taking preferences (for a review see
\cite{lls2}). Such models allowed to uncover and study:
\begin{itemize}
\item Market effects of arbitrarily inhomogenous and non-rational traders behavior.
\item Returns stochastic properties: autocorrelations, volatility,
trading volume
\item Predation, competition and symbiosis among species.
\item Heavy-tailed market returns distributions related to the
ratio between the capital entering the market and the increase
in market stock capitalization.
\end{itemize}

In order to understand what are the crucial factors governing this
complex dynamics displayed by the microscopic representation of
markets, one constructed more schematic models which, while discarding
some of the realistic features of LLS, still conserve the crucial
dynamical features of the market.

The GLV is such a model that embodies some stylized features of the
LLS model in a more generic framework. Instead of following in detail
the way the market price influences each investor population and
individual $i$ , it was assumed that this influence can be represented
through multiplying their wealth $w_i (t)$ by stochastic
multiplicative factors $\lambda_i (t)$. This is naturally suggested by
the LLS model simulations in which the investments of the individuals
(and consequently their returns) are fractions of their wealth (as
implied by the constant relative risk aversion utility functions).
This is also consistent with the recent measurements by \cite {mills}
of the exponent of the power law distribution of the market order
volumes.

The stochastic proportionality between personal returns and personal
wealth is consistent with the real data that show that the (annual)
individual income distribution is proportional to the individual
wealth distribution \cite{pl}.  We proposed
\cite{sorin1,sorin2,sorin3,sorin4,sorin5} therefore a model including
the above stochastic autocatalytic properties of the capital as well
as the cooperative, diffusive and competitive/ predatory interactions
between the investors. The GLV described below is a straightforward
stochastic generalization of the Lotka-Volterra system (and of the
discrete logistic equation) well known previously in population
biology and social sciences.

As explained below, it automatically leads to many of the well known
experimental features of the real markets. However, some of the
initial GLV simplifications were too drastic. In particular the
assumption that the individual returns are extracted from a
probability distribution with fixed variance lead to the result that
the market volatility auto-correlations decays exponentially with
time. This is in stark contradiction with the measured real market
properties \cite{gk,gk1,gk2,gk3,gk4}.  In the present paper we identify the
microscopic dynamical features which are responsible for "volatility
clustering" effect: the fact that the variance of the individual
invested wealth changes is influenced by the global market price
volatility.

We show that this leads also to the emergence of a feedback loop which may in
certain conditions destabilize the market.

\section{Background on the simple GLV model}

More than a hundred years ago, Pareto \cite{pareto} discovered that
the number of individuals with wealth (or incomes) with a certain
value $w$ is proportional to $w^{-1-\alpha}$.

It turns out that in the conditions in which the participants in the
market do not have a systematic advantage one over the other (which is
in fact expected in an efficient market), realistic market dynamics of
the LLS and GLV types lead always to Pareto laws.

Let us define now in more detail the GLV framework: Consider a fixed
constant number of investors (N) (for the extension to a variable
number of agents, see \cite{blank}).  At each time step, the wealth of
each investor ($i$) is $w_i(t)$, and the total wealth is $W(t)=\sum_i
w_i(t)$.

The time evolution of the $w_i (t)$ is simulated by the following
procedure.  At each time step an investor $i$ is chosen randomly to
undergo an event that changes its invested wealth $w_i (t) \rightarrow w_i
(t+1)$.  The various components in the change per unit time $dw_i /
dt$ are:
\begin{itemize}
\item A deterministic component  related to the global status of the economy.
This term is proportional to the current wealth of the investor
through an arbitrary coefficient that may depend on time and on all
the $w$'s:
\beq
m(w_1, . . . , w_N, t) w_i (t)
\eeq
One can imagine that the coefficient $m(w_1, . . . , w_N,t)$
aggregates information on economic growth rate, taxes, social
benefits, interest rates etc. and it is therefore the same for
everybody. This is equivalent to the efficient market hypothesis
\cite{efic}.
\item A purely stochastic term which takes into account the specific
circumstances of each agent. The change of his wealth are still
proportional to its currently invested wealth $ \eta_i (t) w_i(t) $ but the
coefficient $\eta_i $ is a random number taken from a normal
distribution with mean $< \eta > = 0$ and variance $< \eta_i^2 > =
D_i$.
\item There is a social security mechanism, or some fixed relative 
income that ensures that the investors do not become arbitrarily
poor. This term is taken of the form $a_i \sum_j b_j w_j$. The
coefficients $b_j $ represent the relative contribution of the
individual $j$ to the redistributed wealth (through taxes, donations,
payments) while $a_i$ represent the relative amount that the
individual $i$ receives from the redistribution (through salaries,
services, exchanges, pensions, social security). Without loss of
generality one may assume $\sum_j b_j = 1$. In the case in which all
$b_i$'s are constant: $b_i =1/N$, the sum $w (t) = \sum b_j w_j$
reduces to the average wealth $ w(t)= W(t)/N$.
\end{itemize}

Consequently, in the continuum time limit the GLV dynamics is governed
by the system of $N$ coupled non-linear differential stochastic
equations (in th Ito sense) with time dependent coefficients:
\beq
{dw_i \over dt} = [m(w_1 . . . , w_N ,t) + \eta_i (t)] w_i (t) + a_i
\sum_j b_j w_j (t)
\label{richmond}
\eeq
Such systems are notoriously difficult to solve or even characterize
qualitatively.  Yet in the present case, in the limit $N \rightarrow
\infty$, (and for positive, not too unequal $a_i$ and respectively
$b_i$) the probability distribution of relative wealths $w_i / w$ is
completely under analytic control in spite of the fact that the global
wealth is very non-stationary and can have arbitrary ups and downs
(corresponding to booms and crashes/ recession). In particular, with
the notation,
\beq
a = \sum_j b_j a_j (t)
\eeq
the relative wealth:
\beq
  x_i(t) = w_i (t) /w(t)
\eeq
has been shown \cite{pl2,pl3,pl4} to converge even in nonstationary
conditions to a probability distribution that is $ m(w_1, . . . w_N,
t)$-independent:
\beq
\label{powpow}
 P(x_i) \sim  x_i^{-1 -1 -2 a/ D_i}e^{-2a_i /(D_i x_i)} .
\eeq

Consequently, modulo important finite $N$ corrections
\cite{blank} which are outside the present scope, the dynamics
(\ref{richmond}) insures in the range $ x_i > x_{min} \equiv 1/(1+D/a)
$ a power law with
\beq
\alpha = 1+ 2a/D .
\eeq
In the real measurements $\alpha$ has been found to be roughly
constant around 1.5 in the last 100 years in all the western economies
\cite{blank,pl3}.

This value $\alpha \sim 3/2$ has been related to the intrinsic human
biological constraints through the formula $\alpha \sim L/(L-1)$ where
$L$ is the average number of dependents on the average wealth \cite{pl2}.

In the simulations presented here, we will use a discrete version of
Eq. \ref{richmond} with a particular choice of the form of the random
factor and of the parameters: $ m(w_1, . . ., w_N,t) = -a - \mu w(t)$,
$D_i =D$, $a_i = a$, $b_i = 1/N$:
\beq
w_i(t+1)=w_i(t)*e^{\eta_i(t)} +a({W(t) \over N} - w_i(t))- {\mu \over N} w_i(t)W(t)
\label{yor}
\eeq

This specific choice of $m(w,t)$ is the minimal form, which embodies
the 2 main relevant economic facts: 
\begin{itemize}
\item
the term $-a w_i (t)$ represents proportional
taxation, while
\item
the term $-{\mu \over N}W(t)$ models limiting global factors
such as inflation (when the total numerary wealth in the system is
rising, the real value of each investors wealth is decreasing
proportionally.).
\end{itemize}

The simple Eq. \ref{yor} recreates many of the observed features of
stock markets, mainly the observed distribution of the wealth, the
power law in the distribution of market returns and the long term rise
in the total wealth of the investors, it fails to reproduce the long
term correlations in the volatility of the market. We will then show
how to amend this problem and how this influences the market
stability.

\section{Volatility correlations in the simple GLV model}

The main focus of the present paper is the market volatility. Let us
therefore describe first the definition, measurement and properties of
this quantity in the simple GLV model Eq. \ref{yor}. We will then
study in the next section the modified GLV model in which the
volatility determines the variance ($D$) of the random factor $\eta$.

The market return at time $t$ is defined as:
\beq
\label{ret}
r (t) =ln({W(t) \over W(t-1)})
\eeq
The volatility is defined as the average of the square of the returns
over a certain time period (we take it as $N$ unit time steps):

\beq
\label{vol}
V=< [ln(W(t+1)/W(t))]^2>_N
\eeq

The change in W between each time step is small and one can replace eq
\ref{vol} by:

\begin{eqnarray}
\label{main}
V=< [ln({W(t)+\Delta W \over W(t)})]^2> =<[ln(1+{\Delta W \over W(t)})]^2>  \nonumber \\
\sim < [{\Delta W \over W(t)}]^2> = < [(w_i(t+1)-w_i(t))/ W(t)]^2 >= \nonumber \\
= <(x_i(t)*(e^{\eta_i(t)}-1) +a({1 \over N} - x_i(t))- {\mu \over N} x_i(t) W(t))^2>= \nonumber \\
%= <(x_i(t)^2*(e^{\eta_i(t)}-1)^2> +a^2<({1 \over N} - x_i(t))^2> +<{\mu^2 \over N^2} x_i^2(t) W^2(t)> \nonumber \\
%+<x(i)(e^\eta_i(t)-1)(a({1 \over N} - x_i(t))-{\mu \over N} x_i(t) W(t))> -{a \mu \over N}<({1 \over N} - x_i(t)) x_i(t) W(t))> \nonumber \\
%=<x_i^2>D+a^2(<x_i^2>-{1 \over N^2}) +({\mu W(t) \over N})^2 <x_i^2>+{aD \over 2N^2}- \nonumber \\
%{aD \over 2}<x_i^2>-{D\mu W(t)\over 2N} <x_i^2> -{a \mu W(t)\over N^3} +{a \mu W(t)\over N}<x_i^2>\nonumber \\
%=<x_i^2>(D+a^2+({\mu W(t) \over N})^2- {aD \over 2}-{D\mu W(t)\over 2N}+{a\mu W(t) \over N}) -{ a^2\over N^2} +{aD \over 2N^2}-{a \mu W(t) \over N^3}\nonumber \\
%=<x_i^2>(D+a^2+({D \over 2})^2- {aD \over 2}-{D^2\over 4}+{a D \over 2}) -{ a^2\over N^2} +{aD \over 2N^2}-{a D \over 2N^2}\nonumber \\
=<x_i^2>(D+a^2) -{ a^2\over N^2} \sim <x_i^2>D\nonumber \\
\end{eqnarray}

One sees that $V$ depends only on the distribution of the {\bf
relative} wealth $x_i(t)$.

As mentioned above, in GLV the probability distribution of the individual
relative wealth does not change even in the presence of significant
variations in the total wealth. Therefore, the classical GLV, the
volatility is also a stochastic variable with a static
distribution. In particular, the volatility inherits the scaling
properties of the relative wealth distribution. In fact it turns out
that the volatility has a Levy distribution with an index of
$1+{\alpha \over 2}$ (Figure 1).

The experimentally observed long term correlation of the volatility is
a power law i.e: $<{V(t+\tau)V(t)> \over <V(t)^2>} \propto
\tau^{-\delta}$, with an exponent close to $\delta \sim 0$. By
contrast, the time auto-correlation of the volatility decays
exponentially in the simple GLV model.  In order to reproduce the
experimentally observed property we add below an auto-catalytic
dependence of D on V(t).

\section{Dynamic volatility}

The volatility represents the ``nervousness'' of the market. It also
measures the fraction of money an investor can expect to win or loose
by investing in the market during a certain time interval.  Therefore
it is natural to assume that the variance $D$ of the random factor
$\eta(t)$ in Eq. (\ref{yor}) is in fact a function of the volatility
$V(t)$:
\beq
\label{varian}
D=g(V)
\eeq

We will further assume that $g(V)$ can be parameterized by a power of
exponent $n$:
\beq
\label{varian1}
g(V) = c_1 V^n
\eeq

In order to close the feed-back loop one has to estimate the
dependence of the market volatility $V$ on the variance $D$ of the
fluctuations of the individual wealth using equation \ref{main}.  One
can measure $<x^2>$ as a function of D (Figure 2). This dependence turns
out to be linear in the range of values used in the model (0.03-0.1).
Thus in the range of values used in our simulation, one obtains that:
\beq
\label{vola}
V(D)  \approx  c_2 D^2
\eeq
In fact we measured $V(D)$ for the simple GLV model and verified that
it does fit this function (Figure 3).

We are now in the position to estimate the stability of the system as
a function of $n$.  By using the Eqs \ref{varian} and \ref{vola} we
obtain the iterative equation describing the dynamics of the
volatility:
\beq
\label{dyn2}
V(t+1)=c_2 D(t+1)^2 = c_2 c_1 V(t)^{2n}
\eeq
The condition for a stationary dynamics ("fix point") is therefore:

\beq
V= c V^{2n} \Rightarrow V_{fp} = c^{1 \over 1-2n}
\label{stds}
\eeq

where $c=c_1 c_2$

One can now replace Eq \ref{dyn2} with a continuous dynamics:

\begin{eqnarray}
V(t+1) -V(t)= c V(t) (V(t)^{2n-1}-{1 \over c}) \nonumber \\
\Rightarrow \dot{V} = c V(t) (V(t)^{2n-1}-{1 \over c})
\end{eqnarray}

We can now now estimate the stability around the steady state:

\begin{eqnarray}
\dot{\Delta V} = c (V_{fp}+\Delta V) ((V_{fp}+\Delta V)^{2n-1}-{1 \over c
})
\nonumber \\
\Rightarrow \dot{\Delta V}= c (2n-1)V_{fp}^{2n-1}\Delta V \nonumber \\
\Rightarrow \dot{\Delta V}= (2n-1)\Delta V
\end{eqnarray}
The fix point Eq. \ref {stds} is therefore stable for $2n-1 < 0$, and unstable
for $2n-1 >0$. If $n$ is exactly $n=1/2$ then equation \ref{dyn2} will
have a marginal steady state at any value of $V$ if $c=1$.  It will
diverge for any value of $V$ if $c >1$, and it will converge if $c <1
$ to a low enough value of $V$ where the second order dynamics will
take effect.  When the fix point is unstable ($n >1/2$) the dynamics
will lead to zero volatility if we start below the fix point and to
infinity if we start above the fix point.

In order to check this effect we simulated the system 
for various n values: n=1 (figure 4), $n=1/3$ (figure 7,8) and $n=1/2$
(figure 5,6)

\begin{itemize}

\item When $n = 1$ (i.e $2n-1 >0$) The values of V and D both diverge. The
divergence mechanism is not driven by the rise in the wealth of the
investors, or by better investment. It is driven purely by the rise in
the variance of the market. All the feedback interplay takes place
between x (the normalized value of the investors wealth) and the
volatility V(t). The total W(t) does not play an active role in this
feedback loop.

\item When $n=1/3$. The values of the volatility V and of D both
stabilize. This will be the case that will be further analyzed.

\item If $n=1/2$, and we use $c > 1$ the volatility diverges, as expected,
while if $c < 1$ the values of D and V converge to a very low
value. We will further investigate the dynamics that leads to this
steady state.  Interestingly enough, this value at the border between
divergence and stability seems to be favored by the actual market
observations. There might exists an self-organization argument that
explains this fact.

\end{itemize}

We assume the diverging case does not represent realistic situations
though the implied regime might be likened to some of the large
fluctuations experienced in the latest years by the Nasdaq index.

However, we did not consider here, 
the risk-adversity of the investors might be the factor which
ultimately prevents the
unlimited price increase together with the volatility divergence.
We also neglected adverse market effects related with the largest
investors bidding against themselves. 

In the converging cases ($n=1/3$ and $n=1/2, c <1$) long-term
volatility correlations exist (figure 8 and 9). In order to measure
the power of the correlations we made a best-fit estimation to the log
of the correlations with the log of $\tau$ and got for this specific
set of parameters an exponent of $\delta= 0.5$. Note that this
exponent is sensitive to the values of $n$ and $c$. We plan now to
study in a future publication the bounds on the
parameters $n$ and $c$ that can be deduced from the
experimental measurments of the volatility
corelations.

\section{Discussion}
The GLV formalism explains elegantly basic features of wealth
distribution and markets dynamics. One of the main flaws of this
formalism was its failure to predict the long term corelation in the
markets volatility.  Long term correlations imply a long term market
memory. Such a memory is not included in the GLV formalism, which is
based only on the current state of the market. In order for the GLV
formalism to include long term corelations memory must emerge from the
dynamics of the market, instead of being imposed on it externally.

We proposed a simple mechanism that can explain both the long term
corelations of the volatility, and the apparent divergence of some
markets. We propose that the memory is due to a positive feedback
between the volatility and the nervousness of the traders. The
efficient market hypothesis \cite{efic} requires that there can be no
direct link between the average gain of the traders and any of the
measured properties of the market. Thus long term corelations cannot
be due to a positive feedback on the gains. Thus the next natural
candidate for this feedback is the standard deviation of the gains.

The efficiency of the market seems to be also in contradiction with
the fast rise observed in some markets. We therefore propose here that
this rise is not due to the average gain of each investor, but to the
standard deviation of this gain. As a result, one explains the
sustained rise in the prices as the result of the auto-catalytic
feedback of the standard deviation of the traders gain on itself.

We have shown using a stability analysis that a long term rise in the
markets value may take place for certain values of $n$ the exponent
parametrizing the dependence of the nervousness of the traders on the
volatility.

If the trade volume increases sharply as a function of the a rise in
the volatility (very nervous traders) this will lead to a divergent
rise in the market prices.

If on the contrary, the traders are very calm (a weak dependence of
the standard deviation D of the individual returns on the market
volatility V), the markets average returns will be stable, and the
only trace of this dependence will be the long term power-law
auto-corelations of the volatility.

 In a previous paper \cite{pl3} the constance of the individual
 relative wealth Pareto distribution during the last century was
 explained in terms of general sociological factors (size of
 families...) \cite{pl2}. In the present paper we showed that the long
 term evolution of the market may be due to psychological
 factors. Even very weak autocatalytic effects of the volatility on
 the individuals' expectations, can determine the long term evolution
 of the markets.

\newpage
Figure 1 - The volatility distributions.  This distribution has a
power law tail over at least three orders of magnitude. The
distribution was measured for a constant value of $D$ (the variance of
the investors gain), and after the average wealth has stabiliesed due
to the non linear competition factor.  The exponent of the volatility
power distribution is determined by the exponent of the Pareto
relative wealth distribution.

Figure 2 - The average of the square of the relative individual wealth
$<x_i^2>_i$ is found (in the relevant parameters range) to depend
linearly on the variance in the investors random gains $D
=<\eta_i^2>$. Deviations from the linear fit occur:- at very low
values of D, - at large values $ D \sim 1$.  For such values of $D$,
most of the total wealth is in the hands of the wealthiest agent:
$x_{max}=1$ and all other agents have $x_i \sim 0$.  Consequently
$<x_i^2>$ saturates as it approaches its maximal possible value
$1$. However these parameters ranges where linearity is violated, are
outside the interest of the present paper.

Figure 3 - The square root of the volatility as a function of the variance in the gain ($D$), and a linear fit. The gain variance was varied and for each variance we measured the algebric average of the volatility once the total wealth reached equilibrium. Note that the volatility itself has a very large variance, and these results represents the average over a very long time. Thus at small time scales the average volatility may be very different than its long term average.

Figure 4 - The average wealth and volatility, when the variance in the investors gain ($D$) is linearly dependent on the volatility. These represents a very ``nervous'' market in which the einvestors are very sensitive to variations in the stock values. The dynamics in this case diverge through the positive feedback loop between the gain variance and the volatility.

Figure 5 - The average wealth and volatility, when $D$ is proportional to the square root of the volatility. This represents a marginally steady state in which the investors sensitivy to the volatility is precisely inverse to the sensitivity of the volatility on the $D$. The average volatility is stable, however very larfe fluctuations around the average are observed.

Figure 6 - This simulation is simlar to the one presented in figure 5, with a higher value of $c$. The high value of $c$ leads to divergent total wealth and volatility.

Figure 7 - The simulated volatility with the variance proportional to the cubic
root of the volatility. In this case the volatility is stable but it has long term corelations. The two drawings represent the volatility fluctuations at different time scales.

Figure 8 - Long time ($\tau$) auto-correlations of the market
volatility $< V(t+\tau ) V(t) > / < V (t)^2 >$.  The graph was
obtained from runs in which the square standard deviation of the
individual returns D was proportional to $V^{1/3}$. One observes a
straight line fit on the double-logarithmic plot, indicative of a
power-law auto-correlations decay $\tau ^{-\delta}$.

\newpage
\begin{figure}
\center
\noindent
\includegraphics[clip, width = 7 cm]{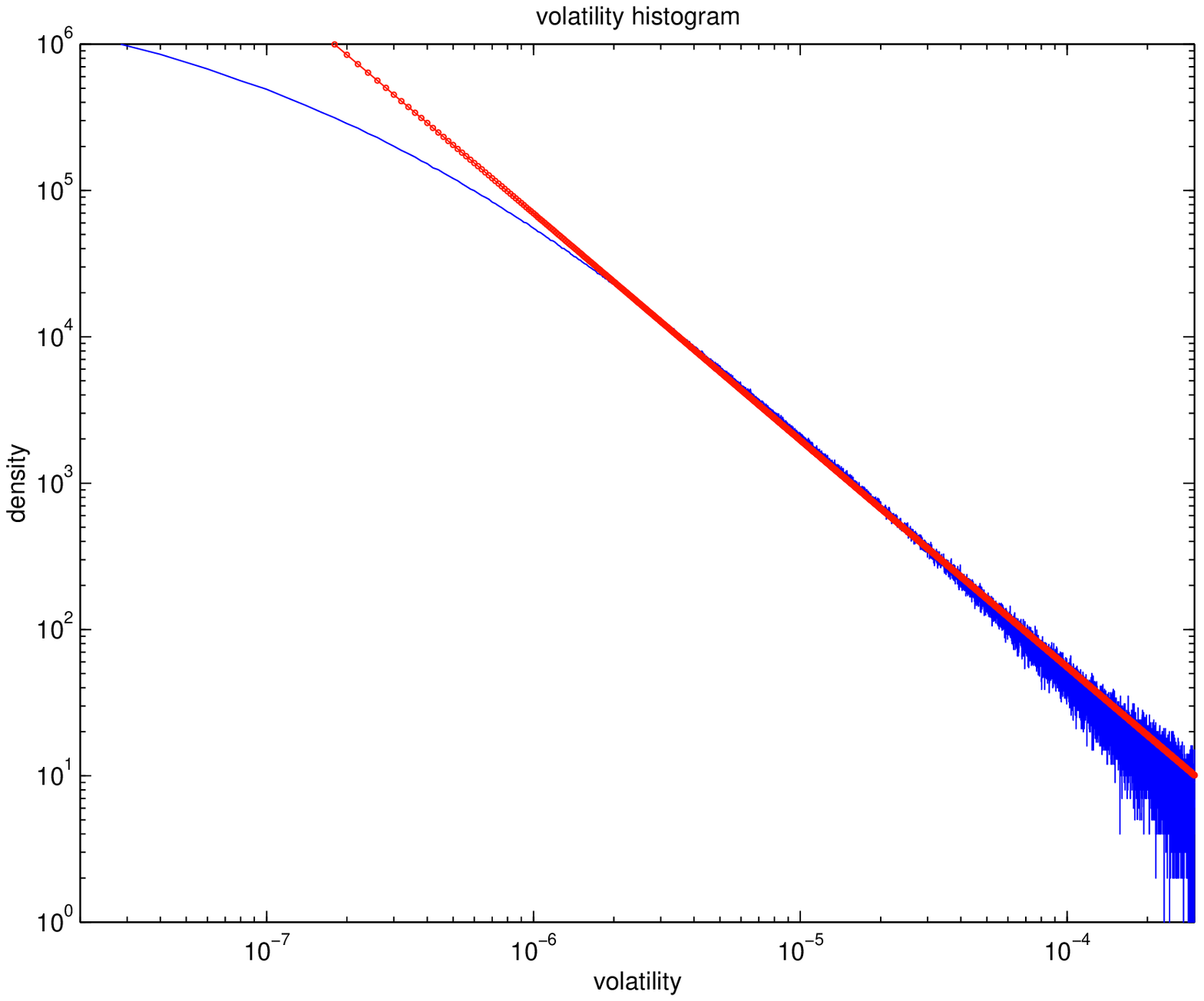}
\caption{}
\end{figure}

\newpage
\begin{figure}
\center
\noindent
\includegraphics[clip, width = 7 cm]{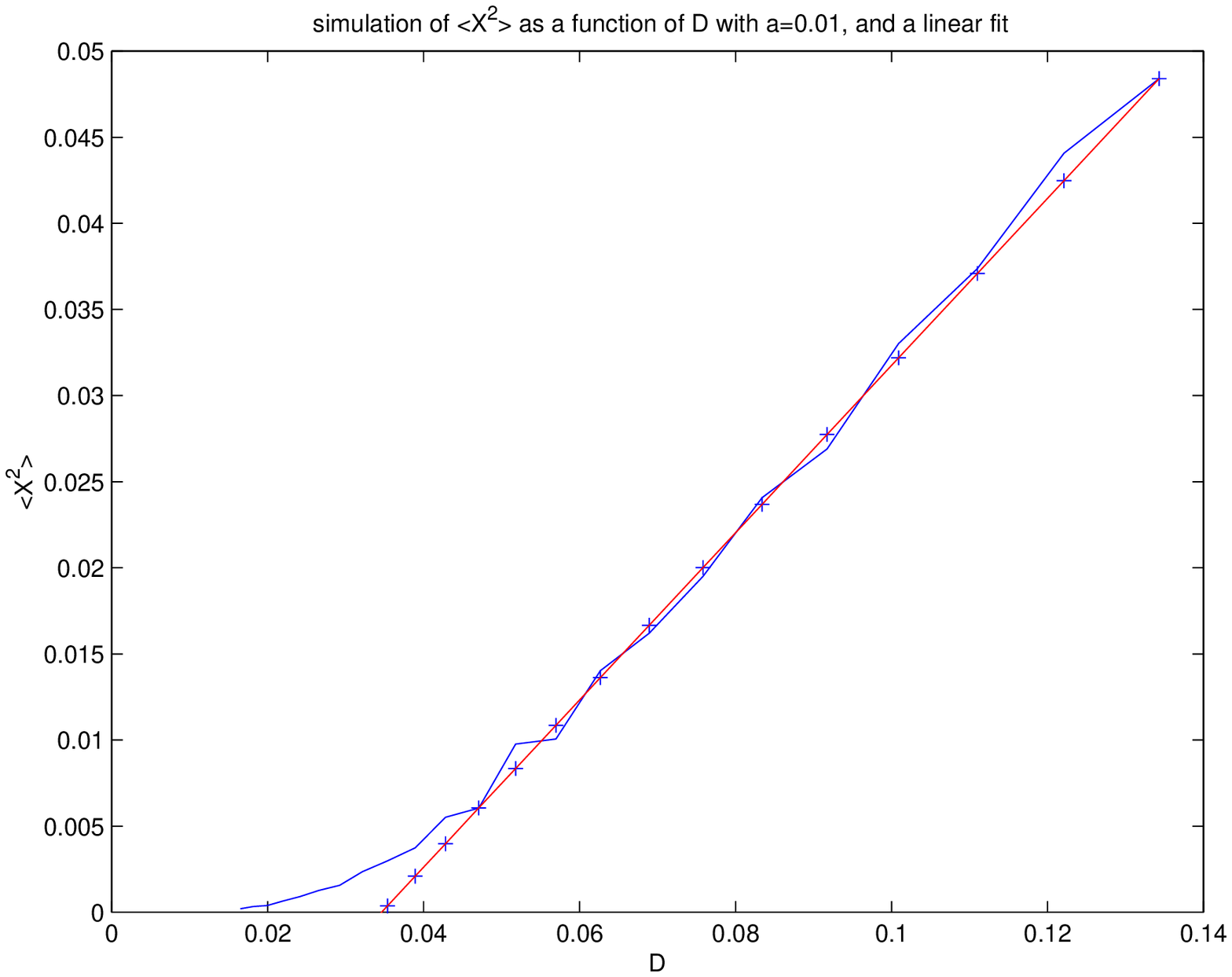}
\caption{}
\end{figure}

\newpage
\begin{figure}
\center
\noindent
\includegraphics[clip, width = 7 cm]{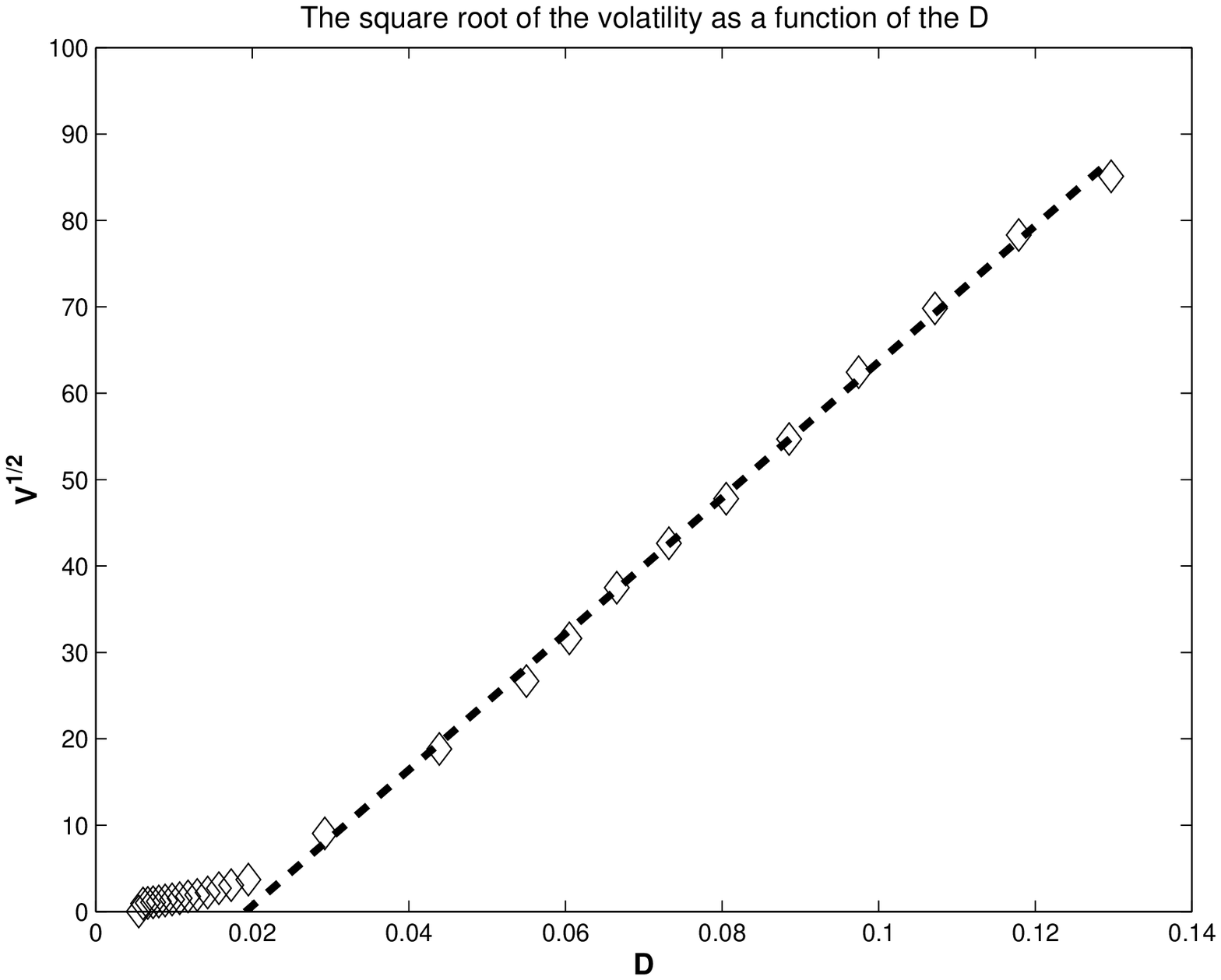}
\caption{}
\end{figure}

\newpage
\begin{figure}
\center
\noindent
\includegraphics[clip, width = 7 cm]{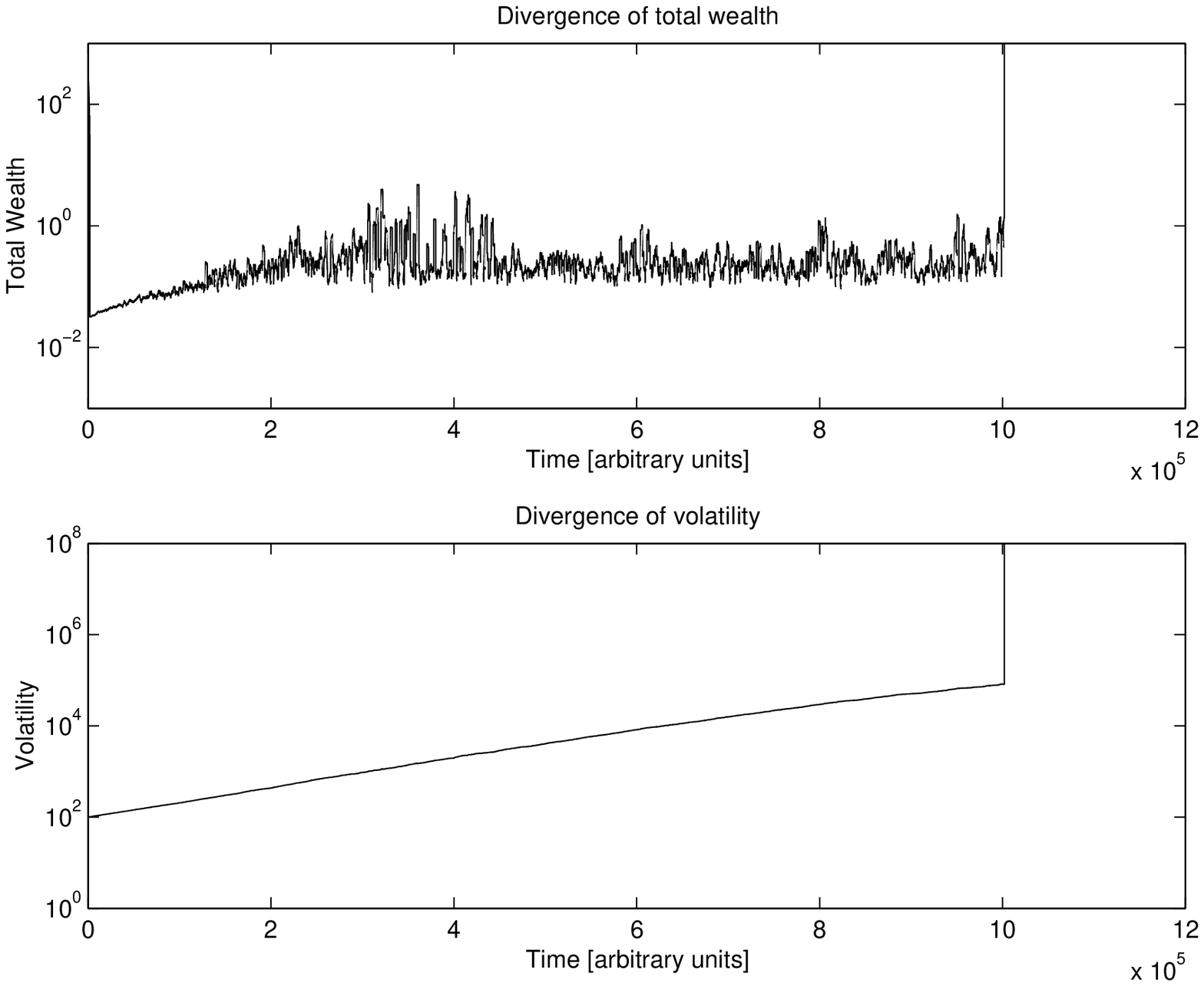}
\caption{}
\end{figure}

\newpage
\begin{figure}
\center
\noindent
\includegraphics[clip, width = 7 cm]{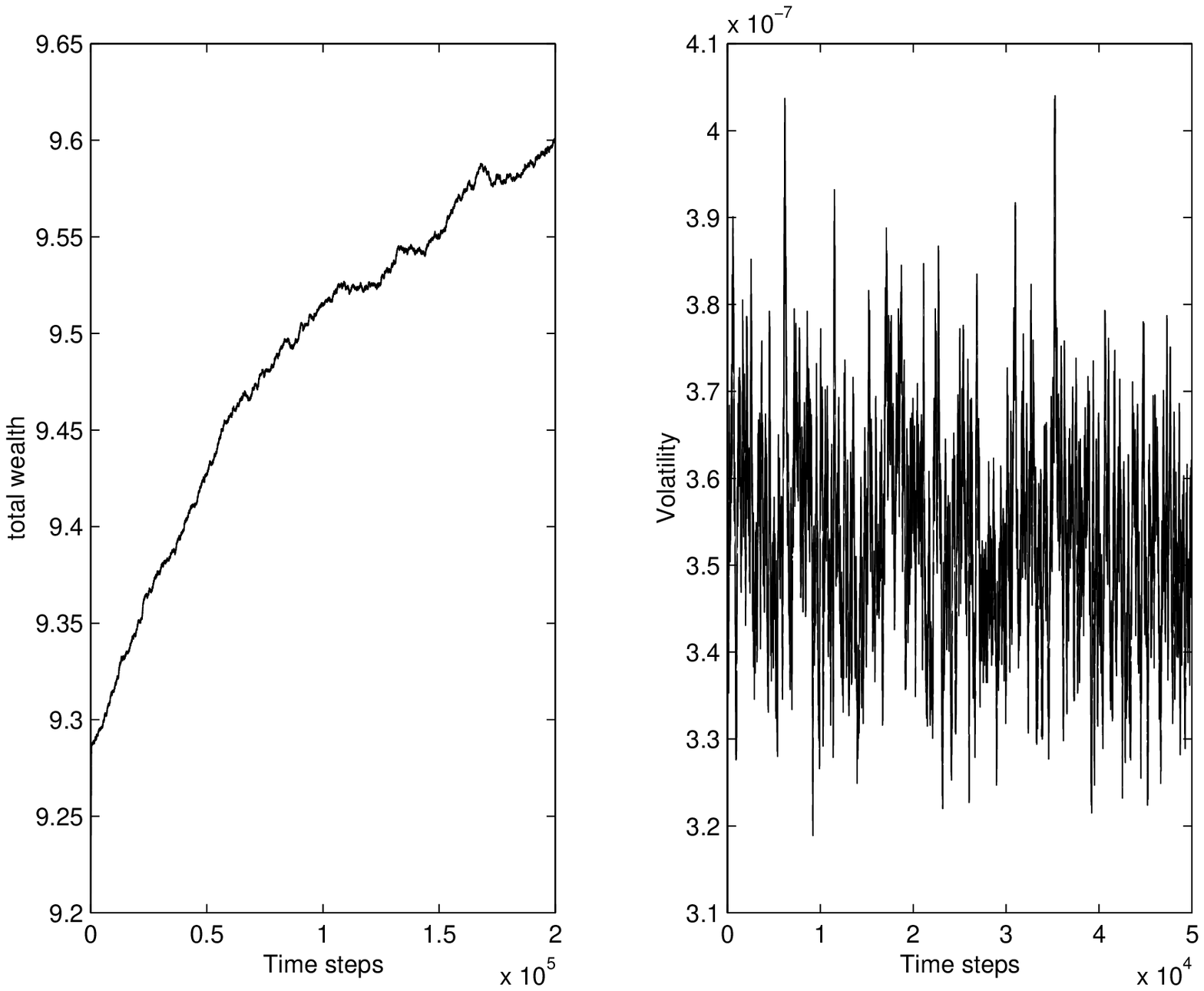}
\caption{}
\end{figure}

\newpage
\begin{figure}
\center
\noindent
\includegraphics[clip, width = 7 cm]{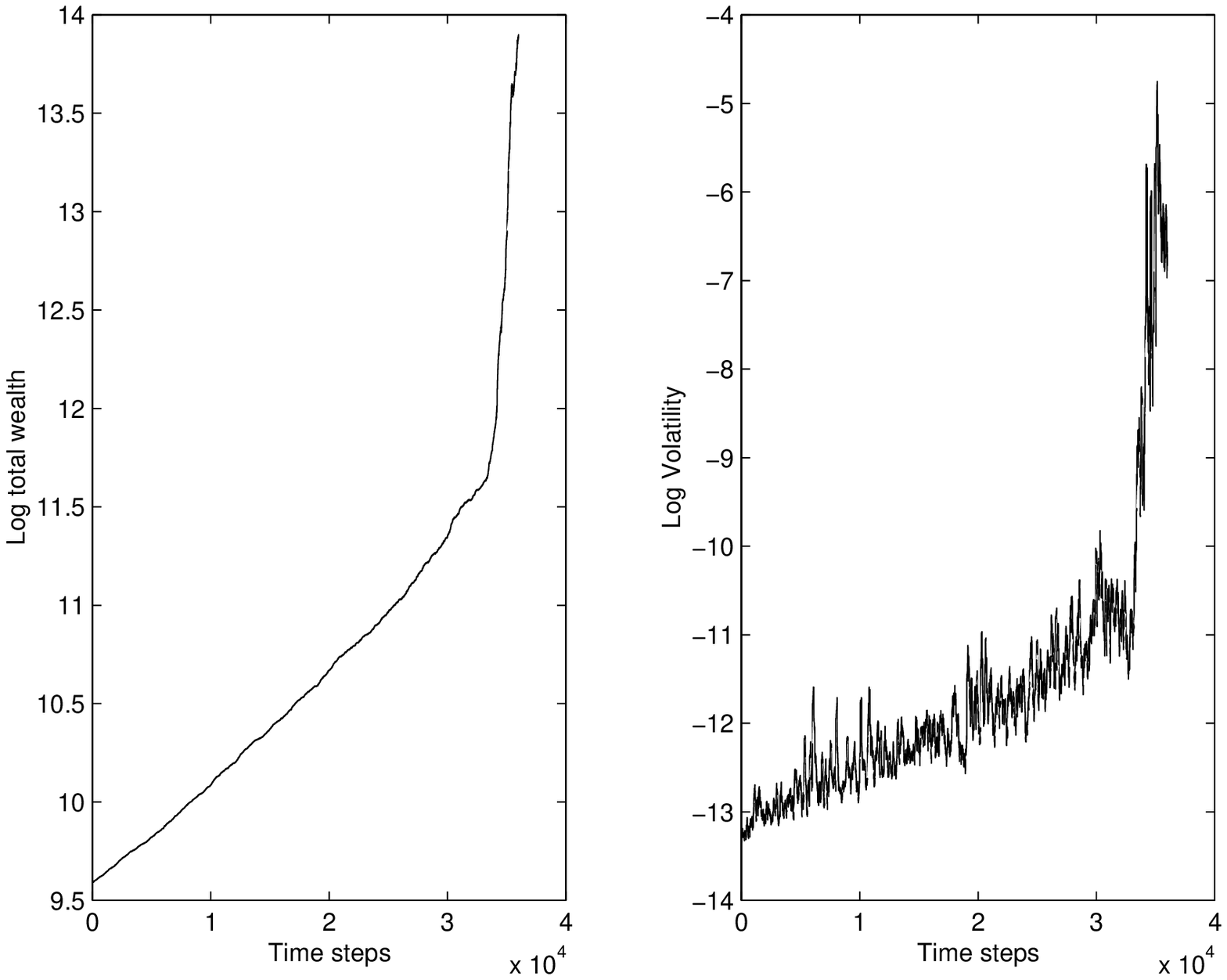}
\caption{}
\end{figure}

\newpage
\begin{figure}
\center
\noindent
\includegraphics[clip, width = 7 cm]{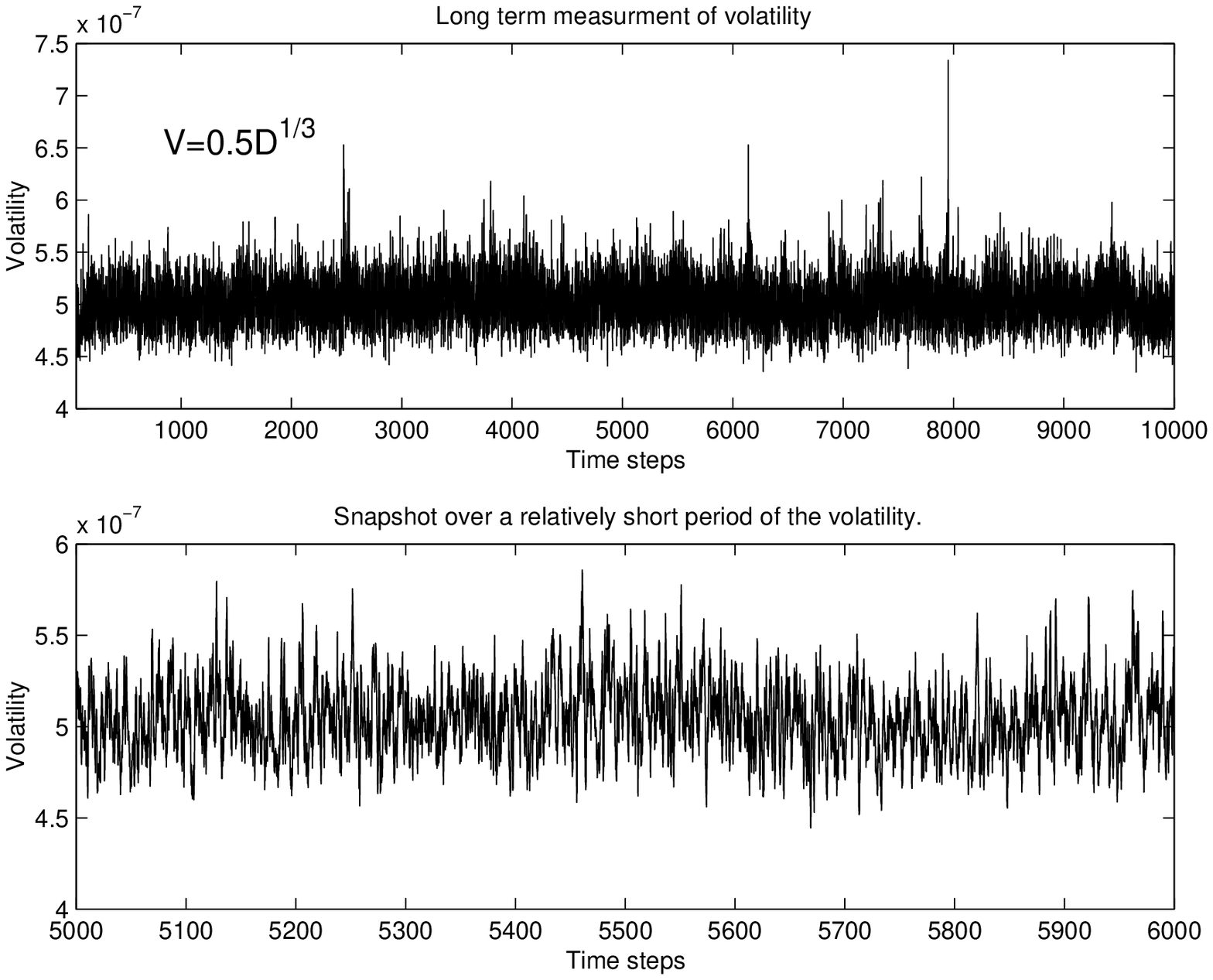}
\caption{}
\end{figure}

\newpage
\begin{figure}
\center
\noindent
\includegraphics[clip, width = 7 cm]{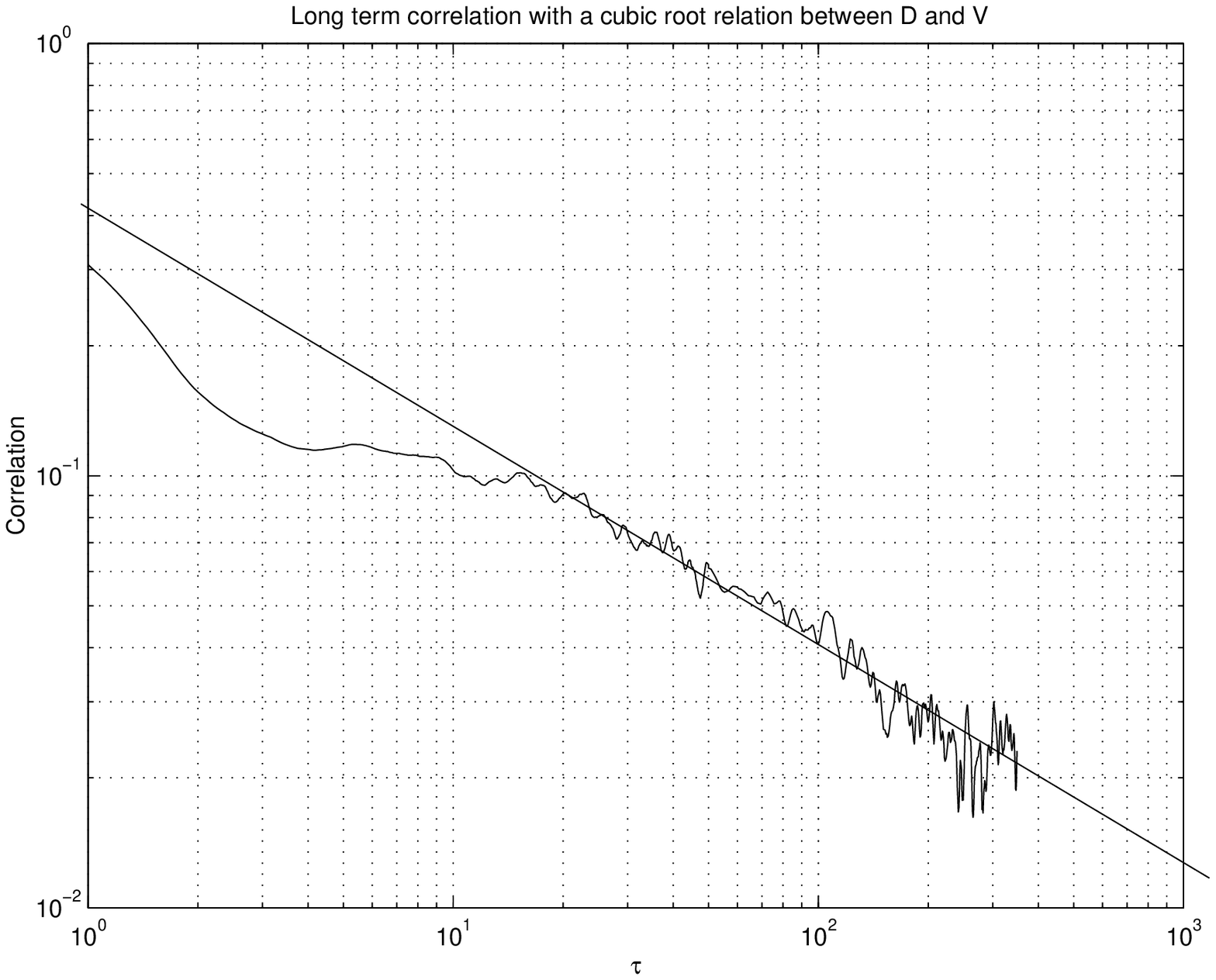}
\caption{}
\end{figure}

\end{document}